\documentclass[journal]{IEEEtran}

\pdfoutput=1

\usepackage{graphicx}
\begin{document}
\title{A Blockchain Consensus Protocol Based on Dedicated Time-Memory-Data Trade-Off}

\author{Miodrag~J.~Mihaljevi\'c% <-this % stops a space
\thanks{Mathematical institute od the Serbian Academy of Scences and Arts, Belgrade, Serbia}% <-this % stops a space
%\thanks{Manuscript received April 19, 2019; revised ...}
}

\maketitle

\begin{abstract}
\textit{A problem of developing the consensus protocols in public blockchain systems which spend a combination of energy and space
resources is addressed. A technique is proposed that provides a flexibility for selection of the energy and space resources which should
be spent by a player participating in the consensus procedure. The technique originates from the cryptographic
time-memory-data trade-off approaches for cryptanalysis. The proposed technique avoids the limitations of Proof-of-Work (PoW)
and Proof-of Space (PoS) which require spending of only energy and space, respectively. Also, it provides a flexibility for
adjusting the resources spending to the system budget. The proposed consensus technique is based on a puzzle where the problem
of inverting one-way function is solved employing a dedicated Time-Memory-Data Trade-Off (TMD-TO) paradigm.
The algorithms of the consensus protocol are proposed which employ certain unconstrained and constrained TMD-TO based inversions.
Security of the proposed technique is considered based on the probability
that the honest pool of nodes generate a longer extension of the blockchain before its update, and a condition
on the employed parameters in order to achieve desired security have been derived. Implementation complexity
of the proposed consensus protocol is discussed and compared with the complexities when PoW and PoS are employed}.
\end{abstract}


% \fi
\IEEEpeerreviewmaketitle

%\IEEEPARstart{T}{his} demo file is intended to serve as a ``starter file''
%for IEEE journal papers produced under \LaTeX\ using
%IEEEtran.cls version 1.8b and later.
%I wish you the best of success.


%\hfill August 26,

\section{Introduction}

In a large number of scenarios we face particular instantiations of the following general problem: All updates of a huge database should be verified before becoming effective. A generic approaches for performing the varication is the centralized one where a trusted party check and verifies all the updates. The main problem with this approach is necessity of exitance a third trusted party as well as the the generic problem of the single point of failure. Recently, as an alternative approach, the blockchain paradigm has been proposed within the bitcoin proposal \cite{bitcoin-Nakamoto-2008}, where the verification is performed in a distributed manner without requirement for the third trusted party as the verification arbiter. The removal of the third trusted party and the distributed verification approach requires an appropriate technique for achieving the verification decision: For this purpose the blockchain based verification employs the so called consensus protocol. This consensus protocol appears as a system overhead. We could say that the escape from the centralized verification paradigm should be paid by the overhead related to the required blockchain consensus protocol.
The overheads implied by the employed consents protocol could be very large and it is an open research issue to construct dedicated
consensus algorithms in order to minimize the overheads in a system based on blockchain technology.

An illustrative simplified framework of a system based on the blockchain paradigm is given in Fig.1

\begin{figure}[h]
	\centering
	\includegraphics[scale=0.35]{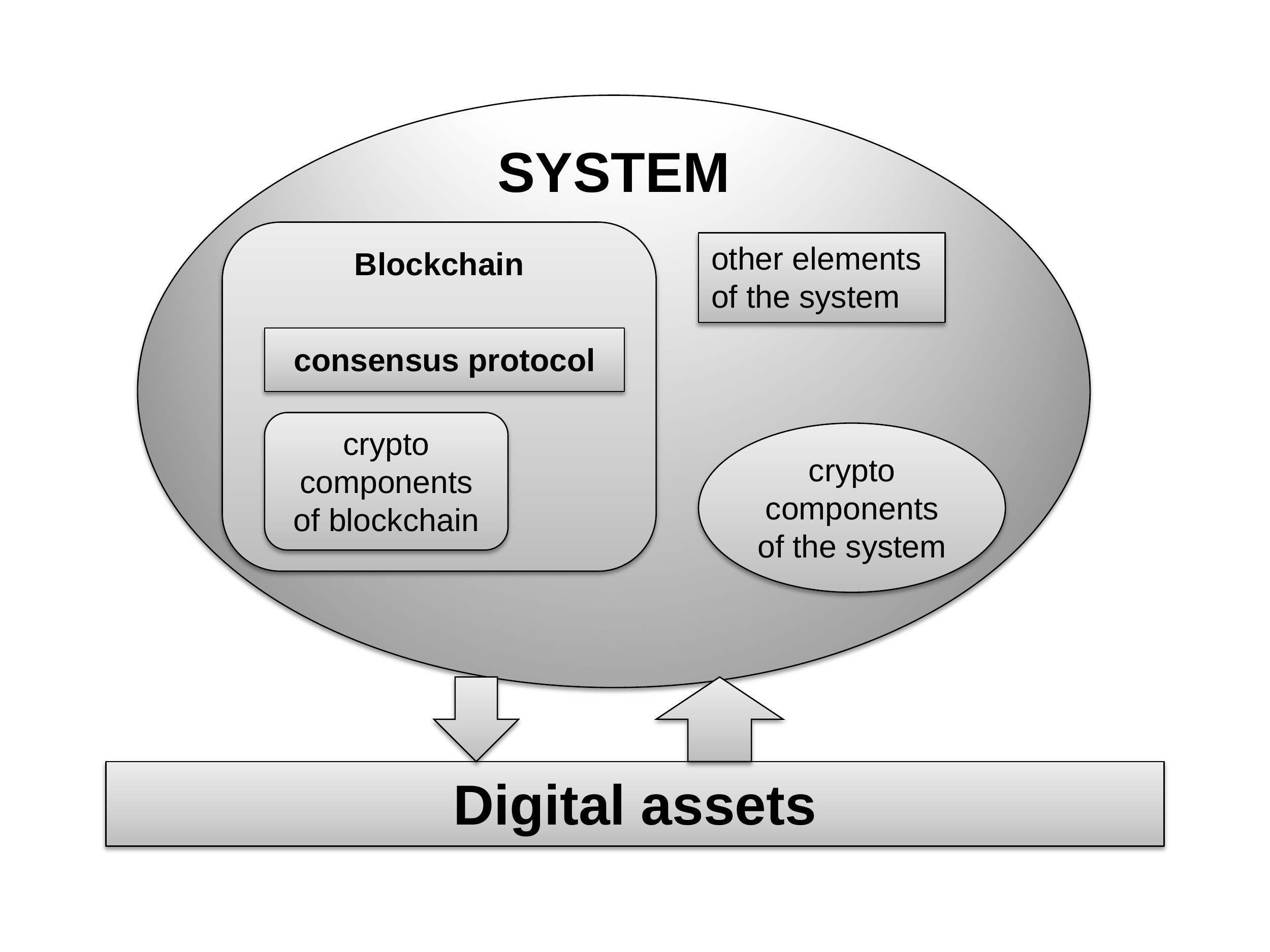}
	\caption{A simplified architecture of a class of blockchain based systems.}
\end{figure}

The consensus protocol is an algorithm which specifies the procedure for generation a candidate block for inclusion into the blockchain, i.e. update/extension of the blockchain with a verified new block.

A lot of consensus protocols for blockchain based system have been proposed and the surveys \cite{Consensus-Review-IEEE_Access-2019-1},
\cite{Consensus-Review-IEEE_Access-2019-2} and \cite{Consensus-Review-IEEE_Comm_Surv-2019}, for example, provide summary of the main approaches.
For the purposes of this paper, we point out to the following two paradigm for achieving the consensus: Proof of Work (PoW) and
Proof of Space (PoS) known also as Proof of Memory (PoM).

PoW approach, initially employed for protection against e-mail spam and denial of service, has been introduced in \cite{Dwork}:
It is a proof approach in which a prover, the party, which claims spending of some resources, convinces a verifier that some computation
with respect to some statement $x$ has been spent.
An illustration of PoW technique is the following: Let $H(\cdot)$ be a hash-function which hashes the concatenation of
given binary sequence $x$ and a randomly selected seed $s$ so that $H(s||x)$ begins with certain number $t$ of zeros.
Assuming that $H(\cdot)$ acts as a random function, the prover must evaluate $H(s||x)$ on $2^t$ different values of $s$ (in expectation)
before ending with a required $s$.
Accordingly, the resource to be spend for PoW approaches is energy.
Nowadays, a family of PoW approaches appear as the most widespread paradigm for securing
the blockchains.

PoS is another paradigm for the consensus protocols - It is based on a proof that certain space/menory has
been booked for participation in the consensus protocol, i.e. the resource which should be spent is "memory".
A simple illustration of the approach is the following:
the verifier specify a random function $f$ which maps a space of dimension $N$ into the same dimensional space.
During the initialization phase, the prover should compute the function table of $f$ and sort it by the output
values. During the verification phase, to convince the verifier that prover posses the table, the prover must
invert $f$ on a random challenge. The previous is just an illustration of PoS underlying paradigm, but this simple PoS
appears as the insecure one.
A secure PoS has been proposed in \cite{pos} employing certain graphs in order to avoid time-memory trade-off approach for
inversion reported in \cite{hellman}. The approach \cite{pos} is such that a cheating prover needs $\Theta (N)$ space or
time after the challenge is known to make the response which the verifier accepts. An application of this PoS has been
employed for developing a crypto-currency reported in \cite{Spacemint-FC2018}.
Another PoS has been proposed in \cite{Proof_of_Space-ASIACRYPT2017} based on inverting random functions from
a class of functions that are hard to evaluate in the forward direction, and even harder to invert.
\cite{Proof_of_Space-ASIACRYPT2017} points out that we only need to be able to compute the entire function table
in time linear (or quasilinear) in the size of the input domain. \cite{Proof_of_Space-ASIACRYPT2017} provides construction of
functions satisfying this relaxed condition and proves lower bounds on time-memory trade-offs beyond the upper bounds given in \cite{hellman}.

\vspace*{0.25cm}
\noindent {\em Motivation for the Work}.
PoW based consensus is based on a puzzle which should be solved employing energy only: The underlying problem has such nature that employment of a memory does not help. On the other hand PoS (PoM) based consensus are built over the problems which could be efficiently solved (solved within given time slots) only if a memory large enough is employed and in this approach particular attention is payed in order to avoid possibility of using time-memory trade-off (i.e. energy-space trade-off) to obtain any benefit. So, a player who participate in the system running (operating), usually called a miner, has no option: The miner must employ a single type of the resources: energy or space - a combination is not allowed. We believe that in a number of scenarios a miner should have opportunity to design the mining budget based on a combination of different resources, and in particular a combination of energy and space in order to deal with solving the consensus puzzle. The flexibility in selection of the resources to be spend
is also important in a context of the incentive required for a miner when the mining should be performed only for a small fee.
On the other hand, we point out that the existing PoW and PoS underlying problems are intentionally selected in order that efficient time-memory or
time-memory-data trade-offs (see \cite{hellman} and \cite{TMD_TO-ASIACRYPT2000}, for example) can not be employed.
Note, PoW and PoS are built on the same paradigms as the cryptanalysis of encryption employing the exhaustive search or
the code-book table, respectively.
Recall that the exhaustive search cryptanalysis assumes search for a solution based on testing all possible candidates for the keys,
and the codebook based cryptanalysis assumes construction of certain table with the pairs (key, ciohertext) which provides inversion
in the look-up manner. On the other hand, time-memory trade-off (TM-TO) and time-memory-data trade-off (TMD-TO)
have been introduced to provide trade-offs of the resources required for the cryptanalysis.
Consequently, our motivation is consideration of design consensus protocols based on the TM-TO (TMD-TO) approaches for trade-off between the resources a miner should employ.

\vspace*{0.25cm}
\noindent {\em Summary of the Results}.
This paper proposes a technique for the consensus in public blockchain systems. The proposed technique provides a flexibility
for selection of the resources which should be spent by a player participating in the consensus procedure. It provides a possibility for
different trade-offs between the required energy and space which should be spent during execution of the consensus protocol.
The technique originates from the cryptographic time-memory-data trade-off approaches for cryptanalysis.
The proposed technique avoids the limitations of PoW and PoS which require spending of only energy and space, respectively.
Also, it provides a flexibility for adjusting the resources spending to the system budget.
The differences of the developed consensus protocol in comparison with the previously reported ones are the puzzle which has
to be solved, construction of the challenge for the puzzle, and the technique for solving the problem. 
The proposed consensus technique is based on a puzzle where the problem of inverting one way function is solved employing dedicated TMD-TO.
Algorithms for the consensus protocol are proposed including dedicated ones for specific inversion problems of one way function where
the challenge is an $\ell$-bits binary prefix (or suffix) $C$ of a ciphertext and the inversion yields one of the keys which encrypt
$n$-bits all ones binary vector into an $n$-bits binary ciphertext with the given prefix $C$, where $\ell \leq n$.
Two options for the inversion problem are considered: the unconstrained one and a constrained one.
Security of the proposed technique is considered based on the probability that the honest nodes generate a longer extension
of the blockchain before its update, and a condition on the employed parameters in order to achieve desired security have been derived.
Implementation complexity of the proposed consensus protocol is discussed and compared with with the complexities when PoW and PoS are employed.

\vspace*{0.25cm}
\noindent {\em Organization of the Paper}.
A framework for the puzzle employed in the consensus protocol is given in Section II. The algorithms of the consensus protocol
when an unconstrained and constrained TMD-TO based inversions are employed are proposed in Sections III and IV, respectively.
Section V yields a security evaluation of the the proposed consensus algorithm. A discussion of implementation complexity of
the proposed approach and comparison with the ones based on PoW and PoS are given in Section VI. Finally, some concluding notes are given
in Section VII.

\section{A Consensus Protocols based on the Puzzle Replacement}

This section proposes a modification of the traditional PoW consensus protocol
where hash-function based puzzle is replaced with a novel one.

\subsection{The Consensus Protocol}

The consensus protocol proposed in this section follows the same framework as the traditional
consensus protocols for public blockchains based on PoW employed in Bitcoin and Ethereum.
Each execution of these protocols consists of the following main phases: (i) Construction a  block of transactions;
(ii) Solving a puzzle; (iii) Inclusion the considered block into the blockchain (assuming that no one transaction from
the block has been already included in the blockchain).
The main differences of the novel blockchain consensus protocol in comparison with the previously reported ones are the puzzle which has
to be solved and construction of the challenge for the puzzle.

\subsection{The Employed Puzzle and the Solving Method}

The puzzle for the consensus protocol proposed in this paper requires, as the first step, construction of a challenge for the puzzle which should be
solved. The challenge is specified as follows: \\
(a) find a nonce, which after added to the binary representation of the block, provides that hash value of the block with the nonce begins with certain number of zeros; \\
(b) take certain number of bits from the hash vector and form the binary challenge vector. \\
The step (a) could employ the same hash-function as employed in Bitcoin and Ethereum consensus protocols and
could be considered as a mini PoW.
The bits selected in the step (b) could be from arbitrary positions of the hash vector, and as a particular instantiation
they could be a suffix (or prefix) of the hash value.
Fig. 2 illustrates the approach employed for construction of the challenge for the puzzle which should be
solved later on.

\begin{figure}[h]
	\centering
	\includegraphics[scale=0.35]{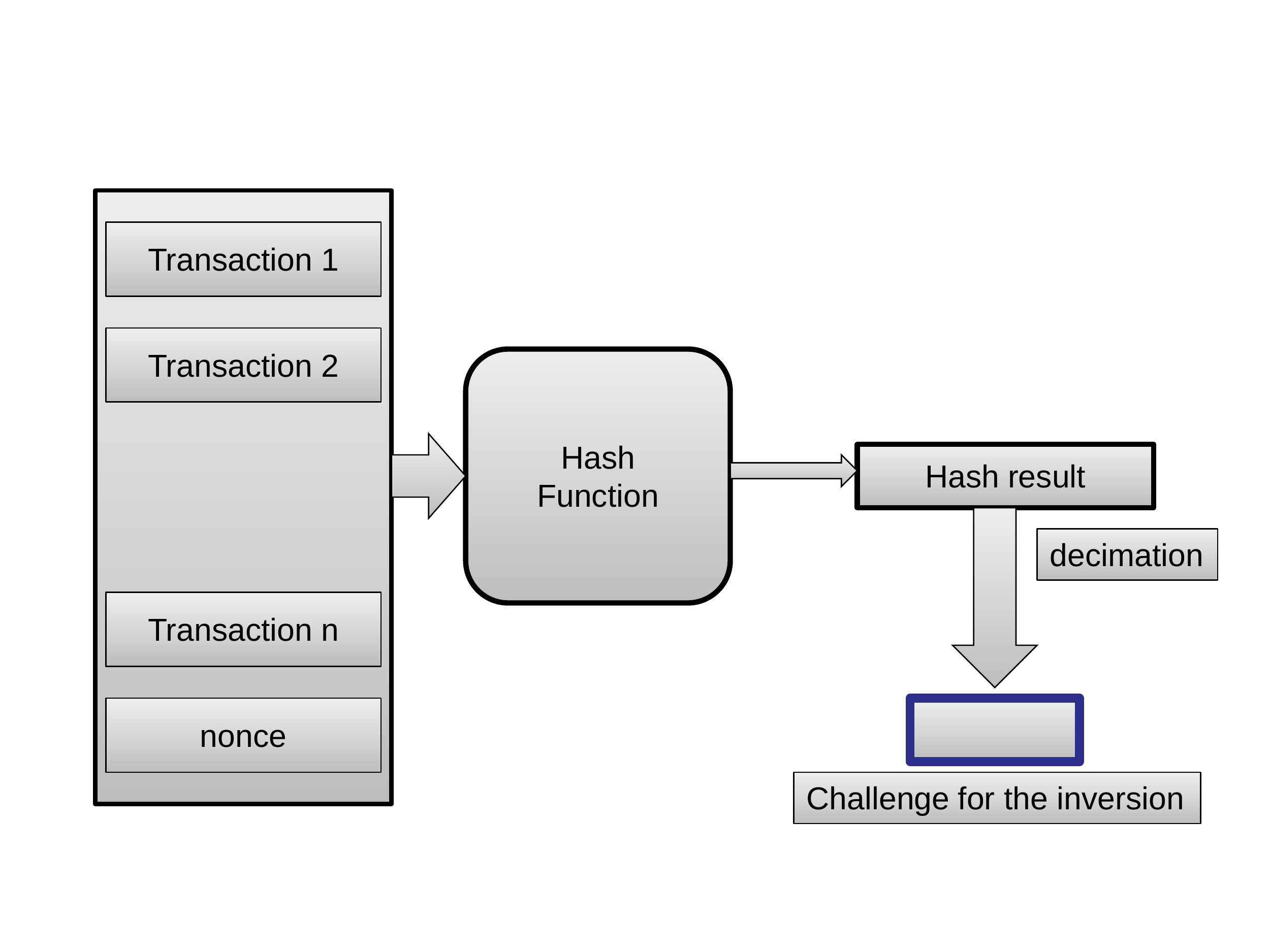}
	\caption{Paradigm of the challenge construction for the puzzle problem.}
\end{figure}

The puzzle itself is an inversion problem for the given challenge: The challenge is considered as
the prefix/suffix of a ciphertxt generated by certain encryption algorithm for the all ones message (or any other given message),
and the puzzle problem is to find a key which provides mapping of the message into the considering ciphertext.
The puzzle paradigm is illustrated in Fig. 3.
Note that this inversion problem could have a lot of solutions because the message is longer than the challenge.

\begin{figure}[h]
	\centering
	\includegraphics[scale=0.35]{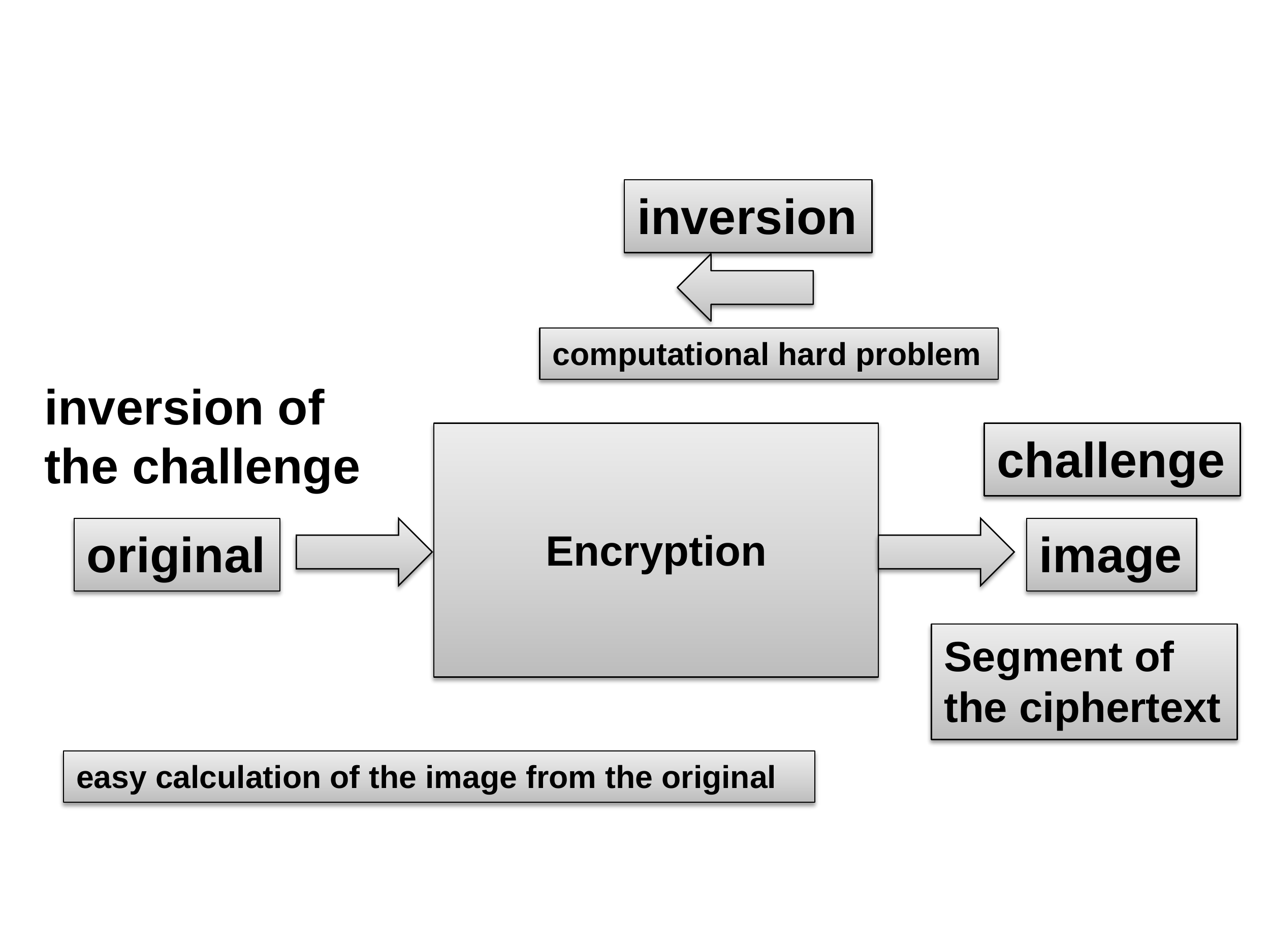}
	\caption{Paradigm of the puzzle.}
\end{figure}

As the method for solving the puzzle, this paper proposes employment of a dedicated time-memory-data trade-off (TMD-TO) approach.
This approach belongs to a family of cryptanalytic approaches which are the generic ones for cryptanalysis of certain
encryption schemes. The proposed approach originate from the TM-TO technique proposed in \cite{hellman}
and later-on generalized in a number of papers. The method for solving the puzzle problem, i.e. the inversion problem, is illustrated in
Figs. 4 and 5, and will be in details explained in the next two sections. This approach follows some underlying ideas reported in
\cite{Mih-IEEE_Comm_Lett-2007} and \cite{Mih-IET_IFS-2012}, as well. The general paradigm of TMD-TO approach is employment of certain
suitably constructed table/tables - these tables should be constructed in advance during
the so called pre-processing phase and later on used for all inversions of the considered class.
Each table consists just two columns: The second column row element is evaluated through a number of recursive recalculations with
the first element of the row as the staring argument as follows:
\begin{displaymath}
[x_{i, j+1}||r_{i, j+1}] = f ([x_{i, j}||r_{i, j}]) \;\;,
\end{displaymath}
\begin{displaymath}
j=0,1,...t-1, \;\, i=1,2,...,m \;\;,
\end{displaymath}
where $[x_{i, j}||r_{i, j}]$ is a binary vector with the prefix $x_{i,j}$ and the suffix $r_{i,j}$,
$f(\cdot)$ is a function under the inversion processing, and $t$, $m$ are the parameters.
In each row of the constructed two-column table the first column element is $[x_{i, 0}||r_{i, 0}]$ and the second one is
$[x_{i, t}||r_{i, t}]$, $i=1,2,...,m$.
An illustration of the table generation is displayed in Fig. 5.

In a general setting, the goal is to find
an argument $x_{i,j}$  such that $[y||b]=f([x_{i,j}||a])$ where $y$ is a given image and $a,b$ are arbitrary.
Consequently, in a general case, we employ a dedicated constrained re-evaluation where the input
for the next iteration step is output of the previous step modified in a deterministic manner.
Note that because, in a general case the given image belong to a subset of all possible images, the inversion results is not a unique one, i.e.
we obtain one of possible arguments which map into the given image.

\begin{figure}[h]
	\centering
	\includegraphics[scale=0.35]{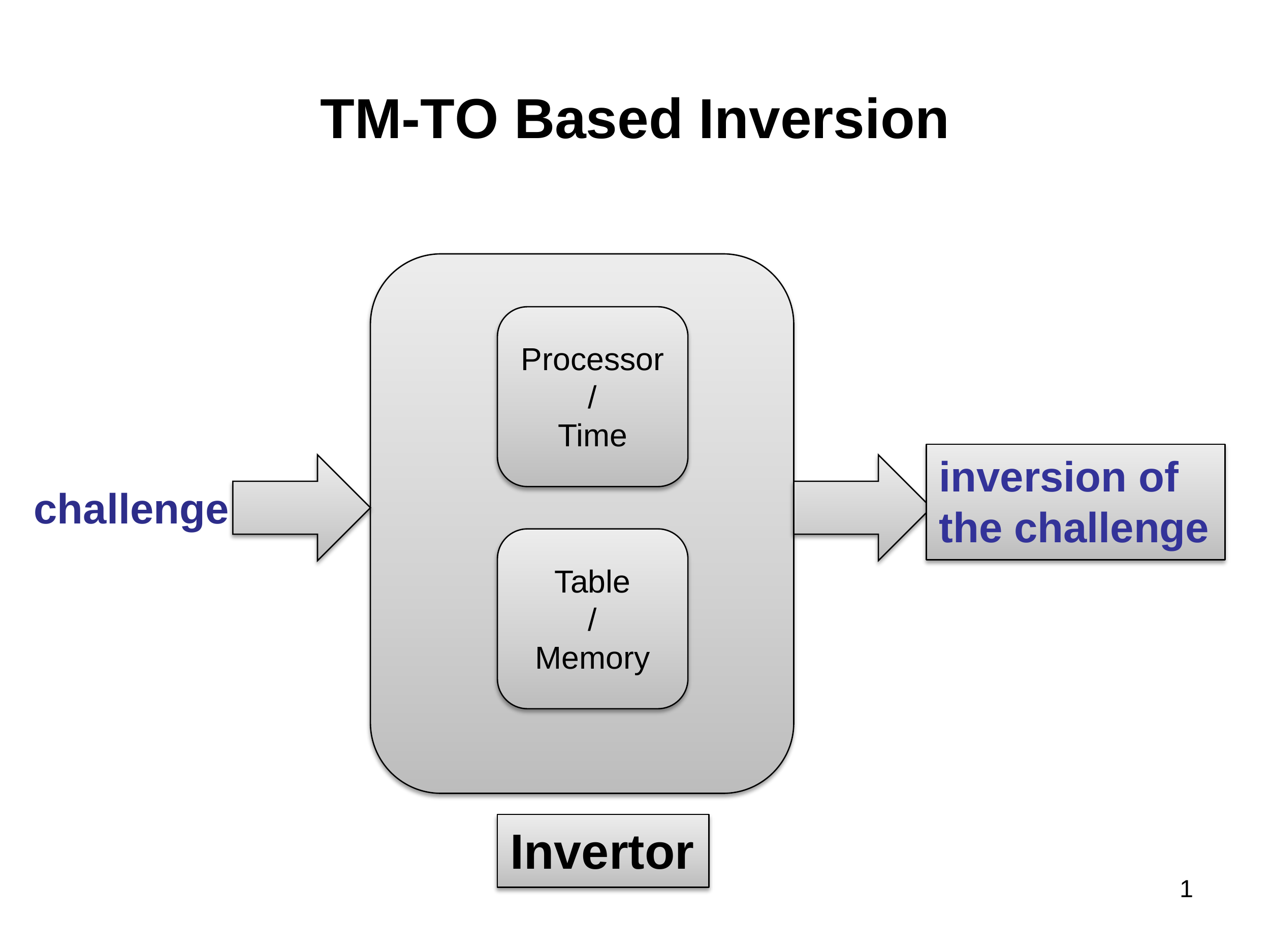}
	\caption{Paradigm for solving the puzzle.}
\end{figure}

\begin{figure}[h]
	\centering
	\includegraphics[scale=0.35]{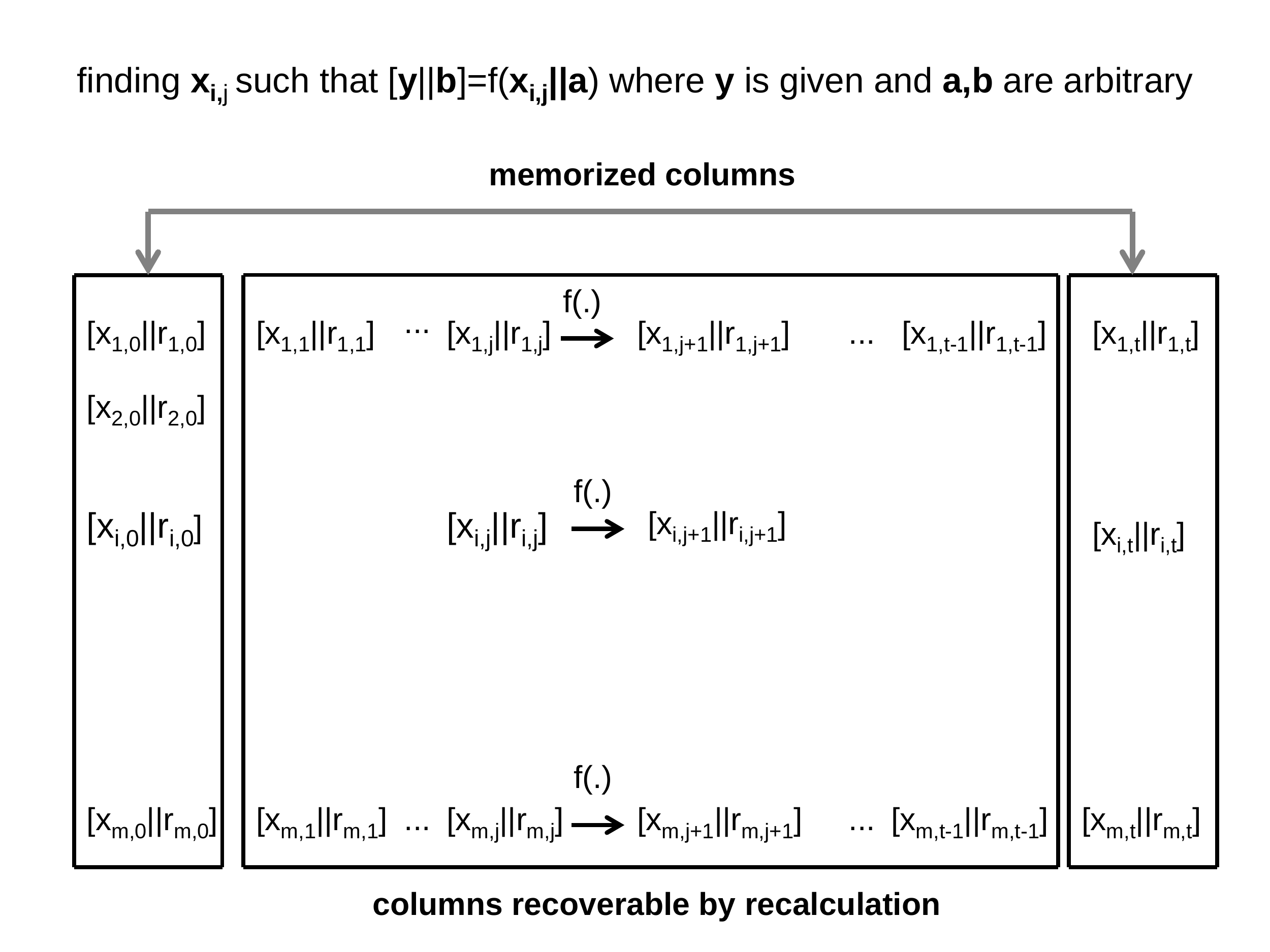}
	\caption{Time-Memory Trade-Off Framework for the inversion.}
\end{figure}

\section{The Basic Approach for the Consensus}

This section yields a basic version of the consensus protocol based on a dedicated TMD-TO.

The consensus protocol for inclusion of a new block into the blockchain, we propose, consists of the preprocessing algorithm and the protocol execution algorithms described in the next two subsections.

For given encryption scheme $E_K(\cdot)$, the inversion problem addressed is the following one: Recover a binary $n$-dimensional key $K$
which encrypts an $n$-dimensional all ones vector $1^n$ into a binary ciphertext with certain $\ell$-dimensional prefix.
Note that in this setting the space of the possible challenges has dimension $2^\ell$ and the inversion result is a point in
the space of dimension $2^n$, implying that the solution is not unique.
Note that, in an alternative setting it is possible to consider inversion problem over a subspace of
the (constrained) inversion results where dimensions of the both spaces are the same and equal to $2^\ell$ - this setting is considered
in the next section.

For simplicity of explanation, we assume that each (elementary) node operates using a single table for each hardness level of the puzzle.
We also assume that the so called federated nodes could exist consisting of multiple independent (elementary) nodes and each of the nodes
employs a single table.

\subsection{Preparation Phase}

The preparation phase should be performed only once and in advance. During this phase certain table should be constructed by
each elementary node.
During the protocol execution this table is used for solving the puzzle.

The inputs for this phase are: (i) $N$ - number of the points in the space of all possible solutions of the inversion problem;
(ii) ${\cal J}$ - set of the parameters which specify the inversion problem hardness, i.e. the puzzle difficulty.

For each level of the hardness/difficulty $j \in {\cal J}$, a node with the index $i$ selects the following two appropriate
(according to the node criteria) space and energy budgets: a memory of dimension $M_i^{j}$ and the processing time parameter $t_i^{j}$.
Note that, in a general case, a node could select that for certain difficulty levels the memory and time budgets are equal to zero.
For the selected parameters, the goal of the preparation phase is to generate certain two columns table with $M_i^{j}$ rows employing
the following Algorithm 1.

\vspace*{0.5cm}
\begin{center}
{\bf Algorithm 1: Initialization of the two-columns table ${\bf M}_i^{j}$}
\end{center}
\begin{enumerate}
\item
{\em Input}: Parameters $n$, $t_i^{j}$, and $M_i^{j}$ (number of the rows in the table ${\bf M}_i^{j}$).
\item
Randomly select $M_i^{j}$ different $n$-dimensional vectors $X_0$ and save them as the first column elements of the table ${\bf M}_i^{j}$.
\item
For each row of the table ${\bf M}_i^{j}$:
\begin{itemize}
\item
perform the following recursive evaluation $t_i^{j}$ times
\begin{displaymath}
X_t = E_{X_{t-1}}(1^n) \,, \;\;\; t=1,...t_i^{j} \;.
\end{displaymath}
\item
Memorize the element $X_{t_i^{j}}$ as the second column element of the table ${\bf M}_i^{j}$.
\end{itemize}
\item
{\em Output}: Two-columns table ${\bf M}_i^{j}$.
\end{enumerate}
\vspace*{0.25cm}

\subsection{The Consensus Protocol}

Proposal of the novel protocol for inclusion of a new block into the blockchain follows the traditional paradigm
which involves construction of a candidate block and solving certain puzzle in order to follow
the consensus achieving approach. The final verification of the blockchain updates assumes that
the longest extension of the previously verified blockchain is accepted as the verified extension for
the further updates.
The protocol we propose contains an alternative puzzle which should be performed over each
candidate block of the updates in order to achieve the verification consensus.
Each node which generates candidate blocks for inclusion into the blockchain should execute the following Algorithm 2.

\vspace*{0.25cm}
\begin{center}
{\bf Algorithm 2: Procedure for Inclusion of a New Block into the Blockchain}
\end{center}
\begin{enumerate}
\item
{\em Input}: New block of the transactions.
\item
Construction of the inversion challenge for considered block of transactions.
\item
Searching for solution of the given inversion challenge; If the solution has been found go to the Step 5.
\item
If execution of the Step 3 does not provide a solution for the inversion, and if processing of the considered block is still relevant go to Step 2; otherwise go to Step 6 (b).
\item
Check the position for inclusion of the considered block into the blockchain after the last already included valid block if no one transaction of the block has already been included in the blockchain; the validity of the predecessor includes correctness check of the inversion problem solution, as well.
\item
{\em Output}: (a) New block inclusion into the blockchain; (b) the failure flag that the block cannot be included into the blockchain.
\end{enumerate}
\vspace*{0.25cm}

Two core components of the above algorithm for inclusion of a new block into the blockchain are:
(i) the algorithm which specify construction of a challenge for the inversion;  and (ii) the algorithm for the inversion search.
Proposals for these two algorithms are as follows.

\vspace*{0.5cm}
\begin{center}
{\bf Algorithm 3: Construction of the Inversion Challenge}
\end{center}
\begin{itemize}
\item
{\em Input}: The considered block of transactions, and the parameters $d$, $\ell^j$
\vspace*{0.15cm}
\item
{\em Construction of the challenge}:
\begin{enumerate}
\item
Randomly select a nonce;
\item
Evaluate the hash value of the binary representation of the considered block concatenated with a nonce;
\item
If the evaluated hash value has $d$-zeros prefix select its $\ell^j$ suffix bits as the inversion challenge and go to the Output; otherwise go to the Step 1.
\end{enumerate}
\item
{\em Output}: Employed nonce and the inversion challenge $C$.
\end{itemize}
\vspace*{0.25cm}

\vspace*{0.25cm}
\begin{center}
{\bf Algorithm 4: Search for a Solution of the Inversion Challenge at the Node $i$ for Given Difficulty Parameter $j$}
\end{center}
\begin{itemize}
\item
{\em Input}: The challenge $C$ for inversion, the difficulty parameter $j$, the table ${\bf M}_i^j$, and the parameter $t_i^j$.
\item
{\em Search for the Challenge Inversion}:
\begin{enumerate}
\item
Check whether the challenge $C$ is equal to $\ell^j$-dimensional suffix of certain second column element of the table ${\bf M}_i^j$; If "yes" record the row where the overlapping has appeared and go to Step 5; otherwise go to Step 3;
\item
Set $t=0$ and ${X}_0 = 1^{n-\ell^j} || C$, i.e., ${X}_0$ is a concatenation of $n-\ell^j$ ones and the challenge $C$;
\item
Set $t = t + 1$; if $t > t_i^j$ go to the Output (b); Otherwise go to the Step 4;
\item
Evaluate
\begin{displaymath}
X_t = E_{X_{t-1}}(1^n)
\end{displaymath}
and check whether $X_t$ is equal to certain second column element of the table ${\bf M}_i^j$; If "yes" record the row index where the overlapping has appeared and go to Step 5; otherwise go to Step 3;
\item
Set $X_0$ to the first column element corresponding to the recorded row index, and perform iterative evaluation  $X_t = E_{X_{t-1}}(1^n)$
until $X_t$ has the suffix $C$, and go to the Output (a);
\end{enumerate}
\item
{\em Output}: (a) Set the inversion result as $X_{t-1}$; (b) the inversion has not been performed.
\end{itemize}
\vspace*{0.25cm}

\section{The Consensus Protocol based on the Constrained Inversion Problem}

This section points out to an alternative version of the previously proposed consensus protocol which corresponds to the setting where
the inversion solutions should be from certain subspace of the entire space of possible solutions.
This setting appears as important one because it provides an additional dimension for controlling
the puzzle difficulty, because the solution should be find within a subset of all possible solutions
for the inversion.
Accordingly, this section proposes the alternative versions of the Algorithms 1, and 4 which correspond
to the consensus protocol based on the constrained inversion problem. \\

\vspace*{0.25cm}
\begin{center}
{\bf Algorithm 1a: Initialization of the two-columns table ${\bf M}_i^{j}$}
\end{center}
\begin{enumerate}
\item
{\em Input}: Parameters $n$, $j$, $\ell^j$, $t_i^{j}$, and $M_i^{j}$ (number of the rows in the table ${\bf M}_i^{j})$.
\item
Randomly select $M_i^{j}$ different $n$-dimensional vectors $X_0$; preset the first first $n-\ell^{j}$ bits of each vector to all ones, and save them as the first column elements of the table ${\bf M}_i^{j}$.
\item
For each row of the table ${\bf M}_i^{j}$:
\begin{itemize}
\item
for $t=1,...t_i^{j}$ perform the following: \\
- if $t=1$ go to the next (evaluate) step, otherwise preset the first $n-\ell^{j}$ bits of $X_{t-1}$ to 1 and go to the evaluate step; \\
- evaluate
\begin{displaymath}
X_t = E_{X_{t-1}}(1^n)  \;\;\;.
\end{displaymath}
\item
Memorize the element $X_{t_i^{j}}$ as the second column element of the table ${\bf M}_i^{j}$.
\end{itemize}
\item
{\em Output}: Two-columns table ${\bf M}_i^{j}$.
\end{enumerate}
\vspace*{0.25cm}

\noindent
Recall that each node $i$ should execute Algorithm 1a for all values of the difficulty parameter $j \in \cal J$.

\vspace*{0.25cm}
\begin{center}
{\bf Algorithm 4a: Search for a Solution of the Inversion Challenge at the Node $i$ for the given Difficulty Parameter $j$}
\end{center}
\begin{itemize}
\item
{\em Input}: The challenge $C$ for inversion, the table ${\bf M}_i^j$, and the parameter $t_i^j$ (corresponding to the difficulty parameter $j$).
\item
{\em Search for the Challenge Inversion}:
\begin{enumerate}
\item
Check whether the challenge $C$ is equal to $\ell^j$-dimensional suffix of certain second column element of the table ${\bf M}_i^j$; If "yes" record the row where the overlapping has appeared and go to Step 5; otherwise go to Step 3;
\item
Set $t=0$ and ${X}_0 = 1^{n-\ell^j} || C$, i.e., ${X}_0$ is a concatenation of $n-\ell^j$ ones and the challenge $C$;
\item
Set $t = t + 1$; if $t > t_i^j$ go to the Output (b); Otherwise go to the Step 4;
\item
for $t=1,...t_i^{j}$ perform the following: \\
- if $t=1$ go to the next (evaluate) step; otherwise preset the first $n-\ell^{j}$ bits of $X_{t-1}$ to 1, and go to the evaluate step; \\
- evaluate
\begin{displaymath}
X_t = E_{X_{t-1}}(1^n)  \;\;\;,
\end{displaymath}
and check whether $X_t$ is equal to certain second column element of the table ${\bf M}_i^j$; If "yes" record the row index where the overlapping has appeared and go to Step 5; otherwise go to Step 3;
\item
Set $X_0$ to the first column element corresponding to the recorded row index, and perform iterative evaluation  $X_t = E_{X_{t-1}}(1^n)$
until $X_t$ has the suffix $C$, and go to the Output (a);
\end{enumerate}
\item
{\em Output}: (a) Set the inversion result as $X_{t-1}$; (b) the inversion has not been performed.
\end{itemize}
\vspace*{0.25cm}

Note that the obtained inversion result should begin with $1^{n-\ell^j}$ prefix, i.e., its prefix should contain a run of all ones of length which corresponds to the considered difficulty.

\section{Security Evaluation}

Let $V_i$ denotes $i$-th verification node and let $\mathcal V$ be the set of all verification nodes. For the security evaluation, we assume that the set $\mathcal V$ consists of two non-overlapping sets: $\mathcal V_H$ and $\mathcal V_M$ corresponding to the
honest and malicious nodes, respectively.  Also, let the chain is updated after each $\Delta$-time slot.
In order to be the node which provides the next block of the blockchain, a node should fulfill the following: (i) face a puzzle solvable by the table in its possession; (ii) be the first of solving the puzzle for a block with transactions which are not included into the blockchain.
In order to be secure, the consensus protocol should be such that the chain version
generated by the nodes from $\mathcal V_H$ is always longer than the chain
generated by the nodes from the set $\mathcal V_M$.

For the security evaluation, without loosing a generality, we assume that each node
is an elementary node which employs a single table: In this model, a system node which employs multiple tables could be considered as a consortium of elementary nodes.
Also, we assume that each table is generated according to the requirements given in \cite{hellman} and \cite{TMD_TO-ASIACRYPT2000}, for example,
so that we expect that all elements of the table in the visible two columns, as well as in the hidden ones are different.
Let $P_i^j$, $i \in \mathcal{V}$, be the probability that a node could solve the puzzle, and
let $P_H^j$ and $P_M^j$ denote the probabilities that a node from the sets $\mathcal V_H$ and $\mathcal V_M$, respectively, could solve the puzzle
of given difficulty $j \in {\mathcal J}$.
%Finally, Let $P^*_H$ and $P^*_M$ denote the probabilities that a node from the sets %$\mathcal V_H$ and $\mathcal V_M$, respectively, be the first to solve the puzzle.

\vspace*{0.25cm}
{\bf Lemma 1.}
Assuming random and mutually independent constructions of the tables for TM-TO corresponding to the given difficulty $j \in {\mathcal J}$, we have the following:
\begin{equation}
P_i^j=\frac{D_i^j M_i^j t_i^j}{N} \;, \; i \in \mathcal{V} \;,
\end{equation}

\begin{equation}
P_J^j=\sum_{i \in \mathcal V_J} \frac{D_i^j M_i^j t_i^j}{N} \;, \; J=H,M \;,
\end{equation}
where $N$ is equal to $2^{\ell^j}$ and $2^{n}$ in the unconstrained and constrained cases respectively, and
$D_i^j$ is the number of employed challenges, i.e. number of the returns from Step 4 to Step 2 during an execution of the Algorithm 2.

\vspace*{0.25cm}
\noindent {\em Proof}.
Assuming the random model, we could expect that $2^{n-\ell^j}$ arguments $X$ fulfils $E_X(1^n) = C$ where we have mapping
$\{0,1\}^n \rightarrow \{0,1\}^{\ell^j}$. Accordingly, the probability that inversion of the given $C$ can be performed by the
given table is equal to $\frac{M_i^j t_i^j}{2^{\ell^j}}$. On the other hand, when we consider the constrained inversion
where the first $n-\ell^j$ bits of the inversion result should be ones, the probability of successful inversion reduces
for the factor $2^{-n+\ell^j}$, i.e. the probability of success becomes $\frac{M_i^j t_i^j}{2^{n}}$. Finally, when there are
$D_i^j$ attempts for the inversion employing $D_i^j$ different challenges, the both probabilities increase for the factor $D_i^j$. \\ Q.E.D.
\vspace*{0.25cm}

\noindent
According to Algorithm 4/4a, note that a node $V_i$ for given difficulty $j \in {\mathcal J}$ requires time $t_i^j$ for each attempt to preform
the requested inversion, and we assume that
$t_{min} \leq t_i^j \leq t_{max}$, $V_i \in {\mathcal V}_J$, $J=H,M$.
Also, we assume that $\delta$ denotes the time required for generation of a challenge in the step 2 of Algorithm 2.

\vspace*{0.25cm}
\noindent
{\bf Theorem 1.}
Assuming that all nodes performs the elementary operations with the same time complexity,
a conservative security condition which provides security of the blockchain
consensus protocol requires the following: For each value of the difficulty
parameter $j \in {\mathcal J}$,

\begin{equation}
\lfloor \frac{\Delta}{\delta + t^{j}_{max}} \rfloor \sum_{i \in \mathcal V_H} {M^{j}_i t^{j}_i} > O(
\lfloor \frac{\Delta}{\delta + t^{j}_{min}} \rfloor \sum_{i \in \mathcal V_M} {M^{j}_i t^{j}_i})\; \;\;.
\end{equation}

\noindent {\em Proof}.
For each $j \in {\mathcal J}$, we have:
\begin{equation}
D^{j}_i = \lfloor \frac{\Delta}{\delta + t^{j}_i} \rfloor \;\;, i \in \mathcal{V} \;\,
\end{equation}
and when $t^{j}_{min} \leq t^{j}_i \leq t_{max}$, $t^{j}_i \in {\mathcal V}_J$, $J=H,M$, for each $i$ and $j$,
\begin{equation}
\frac{\Delta}{\delta + t^{j}_{max}} \leq D^{j}_i \leq \frac{\Delta}{\delta + t^{j}_{min}}
\end{equation}
Consequently, within the time period $\Delta$, each node from $\mathcal V_H$ could consider the puzzle solving for inclusion a block into the blockchain at least $\lfloor \frac{\Delta}{\delta + t^{j}_{max}} \rfloor$ times, and each node from $\mathcal V_M$ could consider the puzzle solving for inclusion a block into the blockchain at most $\lfloor \frac{\Delta}{\delta + t^{j}_{min}} \rfloor$ times.
Accordingly, Lema 1 implies that the expected number of the new blocks included into the blockchain by the nodes from $\mathcal V_H$ is at least:
\begin{equation}
\lfloor \frac{\Delta}{\delta + t^{j}_{max}} \rfloor \sum_{i \in \mathcal V_H} \frac{M^{j}_i t^{j}_i}{N} \;\;, \;\; j \in {\mathcal J} \;\;.
\end{equation}
Also, Lema 1 implies that the expected number of the new blocks included into the blockchain by the nodes from $\mathcal V_M$ is at most:
\begin{equation}
\lfloor \frac{\Delta}{\delta + t^{j}_{min}} \rfloor \sum_{i \in \mathcal V_M} \frac{M^{j}_i t^{j}_i}{N} \;\;, \;\; j \in {\mathcal J} \;\;.
\end{equation}
Consequently, a conservative condition that the number of new blocks added to the chain by the nodes from the set $\mathcal V_H$ is always greater than the number of the new blocks added by the nodes from the set $\mathcal V_M$ can be specified as:
\begin{equation}
\lfloor \frac{\Delta}{\delta + t^{j}_{max}} \rfloor \sum_{i \in \mathcal V_H} \frac{M^{j}_i t^{j}_i}{N} >>
\lfloor \frac{\Delta}{\delta + t^{j}_{min}} \rfloor \sum_{i \in \mathcal V_M} \frac{M^{j}_i t^{j}_i}{N} \; \;\;,
\end{equation}
for each $j \in {\mathcal J}$. The above non-equality can be directly rewritten obtaining a form claimed in the theorem statement. \\
QED.

\section{Implementation Complexity and Comparison}

\subsection{Implementation Complexity}

Implementation complexity of the proposed protocol consists of the following two components: (i) the pre-processing complexity
which is required for the protocol initial setting; and (ii) the processing complexity corresponds to each execution of the protocol.

The pre-processing complexity is complexity of generation the tables employing Algorithm 1/1a, and it is just one-time overhead.

The protocol processing (execution) complexity is determined by time and space complexities of the processing for generation
of a candidate block employing Algorithms 2-4/4a. This processing consists of the complexity of mini-PoW and the complexity
of solving the inversion problem.

Assuming the notations introduced in the previous section, the implementation and execution complexities of the proposed
consensus protocol at a node are summarized in the following table.

\begin{table}[h]
\caption{Time and space complexities of the proposed consensus protocol at an elementary node $i$ for a given difficulty $j$.}
\begin{center}
\begin{tabular}{|c||c|c|}
\hline
                              &                        &                  \\
                              &  time complexity       & space complexity \\
%                              &                        &                  \\
\hline \hline
                              &                        &                  \\
pre-processing complexity     & $O(M_i^j \cdot t_i^j)$ & $O(M_i^j)$       \\
%                              &                        &                  \\
\hline
                              &                        &                  \\
processing complexity         & $O(\frac{\Delta}{\delta + t_i^j} \cdot t_i^j)$ &  $O(M_i^j)$  \\
%                              &                        &                  \\
\hline
\end{tabular}
\end{center}
\end{table}

Consequently, the cumulative energy and space complexities of the protocol within the entire system for adding one block into the blockchain
are as follows:
\begin{displaymath}
O(\sum_i \frac{\Delta}{\delta + t_i^j} \cdot t_i^j))
\end{displaymath}
\begin{displaymath}
O(\sum_j \sum_i M_i^j)
\end{displaymath}
respectively, where the summation is over all elementary nodes $i$ in the system, and the difficulties $j \in {\mathcal J}$.

\subsection{Comparison}

For the comparison, we assume that all the considered puzzles are related to a problem with a solution within a space of $N$ (different) points.
We consider a single node and our goal is to compare time and space complexities of PoW, PoS and the consensus approach proposed in this paper
in order that all three approaches have the same probability $P$ of success in solving the corresponding puzzles.

Note that Algorithm 4 implies that the time complexity of solving the puzzle is $D t$ where $D$ is number of different challenges employed and $t$ corresponds to the time employed for solving the puzzle. On the other hand, according to Lemma 1, the time complexity of the approach proposed in this paper is: $Dt = \frac{N \cdot P}{M}$ where $M$ corresponds to the space complexity of the proposed approach.

\begin{table}[h]
\caption{Comparison of the execution complexities at a node assuming that the the expected probability of success is $P$ and that the complexity of solving the puzzle corresponds to $N$, and $M$ is a parameter $M << N$. }
\begin{center}
\begin{tabular}{|c||c|c|}
\hline
                     &                    &                  \\
                     &  time complexity   & space complexity \\
                     &                    &                  \\
\hline \hline
  PoW                &   $O(N \cdot P)$   &    $O(1)$        \\
\hline
  PoS                &   $O(1)$           & $O(N \cdot P)$   \\
\hline
 proposed approach   & $O(\frac{N}{M} P)$ &    $O(M)$        \\
\hline
\end{tabular}
\end{center}
\end{table}

According to the table, for example, when $M = 2^{30}$: \\
- the energy consumption is reduced about $2^{30}$ times in comparison with PoW requirement at the expense of employing
a memory $O(2^{30})$; \\
- the required space is reduced for a factor $\sim \frac{N \cdot P}{2^{30}}$ times in comparison with PoS requirement at the expense of increase of the required energy for the same factor.

\section{Concluding Notes}

The proposed consensus protocol provides a flexibility for selection of the energy and space resources which should be employed
by a participating entity in the process of verification of the blockchain updates.
Security and complexity of the protocol could be controlled by a number of the parameters: the processing time,
dimension of the memory and the difficulty of the considered puzzle, i.e. hardness of the inversion problem.
These parameters should be adjusted to the particular scenario where the consensus protocol is employed.
Accordingly, a player of the consensus protocol (a miner) could adjust the (mining) strategy in order to fit the expenses into a desired budget.

One of the options for implementation of the proposed consensus protocol is its embedding into the Ethereum platform,
as reported in \cite{Security&CommNtw-2019}. This implementation provides a trade-off between the processing time and dimension of
the employed memory. Illustrative numerical examples of the possible trade-offs between the processing time
and required memory are also pointed out.

An interesting issue for the further work could be a consideration of the time vs. memory issues employing \cite{Proof_of_Space-ASIACRYPT2017}
as a background.

% \appendix[Appendix A]

% or


%\appendices


%\section{}

% \section*{Acknowledgment}

\ifCLASSOPTIONcaptionsoff
  \newpage
\fi

\end{document}